\documentclass[nofootinbib,twocolumn,showpacs,preprintnumbers,amsmath,amssymb,aps,prd]{revtex4}
\usepackage{hyperref}
\usepackage{url}
\usepackage{graphicx}
\usepackage{dcolumn}

\begin{document}
\title{Coating thermal noise for arbitrary-shaped beams} 
\author{R. O'Shaughnessy}
\email{oshaughn@northwestern.edu}
\affiliation{Department of Physics and Astronomy, Northwestern University}
\date{Received ?? Month 2006, printed \today}
\begin{abstract}
Advanced LIGO's sensitivity will be limited by coating
noise.   Though this noise depends on beam shape, and though
nongaussian beams are being seriously considered for advanced LIGO, no
published analysis exists to compare the 
quantitative thermal noise improvement alternate beams offer.    In
this paper, we derive and discuss a simple integral which completely
characterizes the dependence of coating thermal noise on shape.  The
derivation used applies equally well, with minor modifications, to all
other forms of thermal noise in the low-frequency limit.
\end{abstract}
\maketitle

\section{Introduction}
Though gravitational wave detectors such as LIGO are presently taking
data, the best estimates from the astrophysical community for
gravitational waves from compact
object merger rates
\cite{PSconstraints,StarTrack,Chunglee-nsns-1} (though some disagree
\cite{Nakar}), cosmic strings \cite{2006PhRvD..73j5001S}, rotating
neutron stars
\cite{2006PhRvD..73h4001N,2006CQGra..23S...1O,2006PhRvD..73b4021L},
and supernovae \cite{astro-ph..0605493,2006ApJ...640..878B} suggest
that discoveries are most likely to begin with next-generation ground
based interferometers like advanced LIGO.
The present consensus advanced LIGO design has astrophysical reach
(e.g., as measured by the distance to which a pair of inspiralling
neutron stars could be detected) limited by coating thermal noise
\cite{2006ApOpt..45.1569H}.
In this context, thermal noise denotes the phase noise in the IFO
produced by elastic oscillations of the mirror
excited by the thermal bath of the remaining degrees of freedom
\cite{1999PhLA..264....1B}; coating thermal noise denotes strong
contributions to the noise
arising when couplings between elastic modes and the thermal
bath (i.e., losses) are predominantly located in the thin mirror
coating off which the test beam reflects.  
Thermal noise depends
strongly on beam shape: as one can show by applying the
fluctuation-dissipation theorem to a low-frequency limit \citet[see,
e.g.,][]{2000PhRvD..62l2002L}, for gaussian beams thermal noise power
goes as $\propto
r_o^{-2}$.
A flatter beam which more equitably averages over fluctuations, such
as ``mesa'' beams \cite{DOSTV-2004,OSV-2004} or hyperboloidal generalization
\cite{2004.gr-qc..0409083,2004.gr-qc..0409084,2006PhRvD..73l7101G}, should reduce coating
thermal noise.

In this paper, we provide a simple two-dimensional integral to allow
comparison of the coating thermal noise for different mode shapes.  
Specifically, the fluctuation dissipation theorem, plus symmetry arguments about
half-infinite mirrors, plus some scaling arguments imply the
 power spectrum $S_x(\omega)$ of coating thermal noise must be
proportional to the coating thickness $d$ and to two-dimensional
integral over the fourier transform $\tilde{P}(K)$ normalized beam shape $P(r)$
\begin{equation}
 S \propto  d \int d^2 K |\tilde{P}(K)|^2
\end{equation}
where  $\tilde{P}(K)$, its two-dimensional fourier
transform. 

However, the symmetry arguments presented can be applied to nearly any
system with (approximate) two-dimensional translation symmetry (i.e.,
with a small beam on a large mirror): 
$S\propto \int d^2 K K^p |\tilde{P}(K)|^2$ for some constant
index $p$.  This index can be uniquely determined by comparison to
other calculations for gaussian beams; thus the correct noise
dependence on beam shape can be easily determined and understood for  bulk
thermoelastic noise ($S\propto r_o^{-3}$ implies $p=1$) \cite{OSV-2004,1999PhLA..264....1B}, for bulk
thermal noise ($S\propto r_o^{-1}$ implies $p=-1$) \cite{1998PhRvD..57..659L,1999PhLA..264....1B,1998PhRvD..57..659L},
and for coating thermoelastic noise ($p=0$) \cite{2004PhRvD..70h2003F}.

\section{Scaling argument}
According to the fluctuation-dissipation theorem, coating thermal noise is proportional 
to the power dissipation rate $W_\text{diss}$ associated with a fluctuating 
pressure of shape $P(r)$ on the mirror surface.  Manifestly (for half-infinite mirrors), $W_\text{diss}$ must be proportional to a translation-invariant inner product on $P$, of form
\begin{eqnarray}
  W_\text{diss}\propto \int d^2 R \int d^2 R' V(R-R') P(R) P(R') \nonumber\\
 \propto \int d^2 K \tilde{G}(K) |\tilde{P}(K)|^2
\end{eqnarray}
By definition, the coating
thermal noise is the contribution of the coating to the total thermal
noise; thus, expanding in powers of coating thickness,
\begin{equation}
\tilde{G}(K,d) \approx \tilde{G}_o(K) + d \tilde{G}_1(K) + \ldots
\end{equation}
Since no other transverse scale exists in the half-infinite
mirror, the kernel $\tilde{G}_1$ must be scale invariant, and
therefore satisfy $\tilde{G}_1(\lambda K)=\lambda^p \tilde{G}(K)$, and thereforem be
of form 
\begin{equation}
\tilde{G}_1(K) = K^p c_1
\end{equation}
for some constant $c_1$.  Finally, to recover the usual result for
gaussian beams  
(i.e. $S\propto d/r_o^2$, as has been extensively calculated
\cite{2006ApOpt..45.1569H,2002PhRvD..65j2001N,1998PhRvD..57..659L}), we must have  $p=0$.

\section{Detailed calculation}
To check this simple scaling argument, we can perform the full
fluctuation-dissipation calculation of coating thermal noise  in a
special case where the exact solution is known: when the elastic
properties of the medium and coating are identical.  The relevant
elastic green's functions for a half-infinite mirror are
provided in an appendix of \citet{2002PhRvD..65j2001N}. 
From Nakagawa et al's Eq. (1) , we know 
\begin{equation}
  S \propto \int d^2 R \int d^2 R' P(R) P(R') \text{Im}\chi_{zz}(R-R')
\end{equation}
where $\text{Im}\chi_{zz}(R,R')$ is given by their Eqs. (4-5):
\begin{eqnarray}
\text{Im}\chi_{zz}(r) 
&=&
 \phi \frac{1-\sigma^2}{\pi E} \left[ F(r,0)-F(r,d) \right] \\
F(r,z)&=& \frac{1}{\sqrt{r^2+4z^2}} \nonumber \\
 & & \times \left(
  1+ \frac{z^2/(1-\sigma)}{r^2+4z^2} 
   + 12 \frac{z^4/(1-\sigma)}{(r^2+4z^2)^2}
  \right)
\end{eqnarray}
We can equivalently represent this integral in the fourier domain, as 
\begin{equation}
  S \propto \int d^2 K |\tilde P(K)|^2 \left[ \tilde{F}(K,0) -\tilde{F}(K,d) \right]
\end{equation}
where [Nakagawa et al Eq. A1]
\begin{equation}
  \tilde{F}(K,d) = 2 \pi \frac{e^{-2 K d}}{K} 
  \left[
    1+ \frac{K d}{1-\sigma} + \frac{(K d)^2}{1-\sigma} 
  \right]
\end{equation}
In other words
\begin{eqnarray}
S&\propto& \int_0^\infty  d^2K  |\tilde P(K)|^2 \\
 & & \times \left[
 \frac{-1+\exp[-2 Kd]}{K} + \frac{d \exp[-2 K d] (1+Kd)}{1-\sigma}
  \right] \nonumber
\end{eqnarray}
Naturally, $\tilde{P}(K)$ drops to zero well before $K\approx 1/d$; therefore, we may take a small-$d$ limit.  We therefore conclude 
\begin{equation}
S \propto \int d^2 K
|\tilde{P}(K)|^2 \; .
\end{equation}

\section{Conclusions}
Complementing similar earlier studies by
\citet{OSV-2004} on thermoelastic noise. in this paper we describe how to
calculate how coating thermal noise varies  with beam shape.  An
independent derivation, as well as detailed discussion of alternative
beam-shape applications,  will be forthcoming by other authors
(Lovelace et al, in preparation).

More generally, this paper describes a simple way to unify several
disparate calculations for the beam shape dependence of
thermally-driven noise (e.g., coating thermal noise; bulk
thermoelastic noise; bulk thermal noise) produced when a beam reflects
off a large mirror. 



\begin{acknowledgements}
Geoffrey Lovelace deserves considerable thanks for pointing out an
error (an incorrect power index for coating thermoelastic noise).  
This work is partially supported by NSF Gravitational
Physics PHYS  grant-0353111.
\end{acknowledgements}

\bibliographystyle{astroads}
\bibliography{apj-jour,LIGO-design-2-mirrors-coatings,LIGO-design-2-mirrors-control,popsyn,short-grb-data-analysis,supernovae-theory,gw-astronomy,gw-astronomy-pulsars,gw-astronomy-mergers}

\end{document}